\documentclass{pasa}
\usepackage{float}
\usepackage{color}
\usepackage{multicol, blindtext}
\usepackage{url} 
\usepackage{hyperref} 
\usepackage{breakurl}
\usepackage{amsmath}
\usepackage{graphicx} 

\title[Gaia TRGB]{Galactic Calibration of the Tip of the Red Giant Branch}

\author[Mould et al.]{{Jeremy Mould$^{1}$, Gisella Clementini$^2$ and Gary Da Costa$^3$}\\
\affil{$^1$Swinburne University, PO Box 218, Hawthorn, Vic 3122, Australia}
\affil{$^2$INAF - Osservatorio di Astrofisica e Scienza dello Spazio, 40129 Bologna, Italy}
\affil{$^3$Research School of Astronomy \& Astrophysics, Australian National University, ACT 0200,
 Australia}}

\jid{pasa}
\doi{10.1017/pas.\the\year.\#}
\jyear{\the\year}

\hypersetup{colorlinks,citecolor=blue,linkcolor=blue,urlcolor=blue}

\begin{document}
\begin{abstract}
Indications from Gaia data release 2 (DR2) are that the tip of the red giant branch (TRGB, a population II
standard candle related to the helium flash in low mass stars) is close to --4
in absolute I magnitude in the Cousins photometric system. Our sample is high latitude
southern stars from the thick disk and inner halo, 
and our result is consistent with longstanding findings
from globular clusters, whose distances were calibrated with RR Lyrae stars.
As the Gaia mission proceeds, there is every reason to think an accurate Galactic 
geometric calibration of TRGB will be a significant outcome
for the extragalactic distance scale.
\end{abstract}
\begin{keywords}
Parallax, Stars -- red giants, galaxies: distances and redshifts 
\end{keywords}

\maketitle
\section{ Introduction}
The goal of 1\% accuracy in galaxy distances is now driven more by questions in fundamental physics than astronomy. A roadmap to reach this goal exists by means of observing and modelling cosmic microwave background anisotropies (Di Valentino et al 2018). The astronomical distance ladder also has a path to reach this goal by calibrating the Cepheid period luminosity relation and the type Ia supernova standard candle. A second population II approach exists in Tip of the Red Giant Branch (TRGB) distance measurements (Mould 2017; Beaton et al 2016).

Da Costa \& Armandroff (1990) first showed using Milky Way globular clusters that the magnitude of the tip of the red giant branch (TRGB), which is set by the luminosity of the He core flash, can be used as a distance indicator for old stellar populations.  Subsequent work (e.g., Hatt et al., 2018 and references therein) have demonstrated the utility of the TRGB for distance determinations.  Here we explore an alternative means of calibrating the TRGB method by using Gaia DR2 parallaxes for field red giants that lie at high Galactic latitude (Gaia collaboration et al.\ 2018b). Clementini et al.\ (2018) 
present the processing and catalogues of RR Lyrae stars and Cepheids released in Gaia DR2. 

\section{Red Giant Sample}
The ideal sample for this purpose is a stellar population drawn from the thick disk and the inner halo of the Milky Way.  The density profile of the halo follows a power-law with an index of order --3 out to a radius R $\approx$ 25--30 kpc (e.g., Saha 1985; Iorio et al. 2018, and references therein) but has a steeper fall-off beyond this radius (e.g., Hernitschek et al., 2018, and references therein).  This means there is little to be gained by pursuing TRGB stars to great distances, where the 
relative errors are larger.

\subsection{Database Query}
We use as our input catalogue Data Release DR1.1 of the SkyMapper survey of the southern sky (Wolf et al. 2018) which incorporates a crossmatch to Gaia DR2 catalogue and selected stars fainter than a SkyMapper $i$ magnitude of 9, where saturation effects begin to appear, and brighter than Gaia $G$ magnitude of 14 so that the Gaia image centroiding is excellent.  The specific database query was:

{\it \noindent select top 250000 raj2000, dej2000,}\\
{\it i$\_$psf, e$\_$i$\_$psf, i$\_$nimaflags, bp$\_$rp,\\
parallax, parallax$\_$error, z$\_$psf, astrometric$\_$excess$\_$noise\\}
{\it FROM DR1.master m join ext.gaia$\_$dr2 on \\
(source$\_$id=m.gaia$\_$dr2$\_$id1 and }\\
{\it m.gaia$\_$dr2$\_$dist1$<$1) WHERE (glat$<$=-36) AND \\
(phot$\_$g$\_$mean$\_$mag$<$=14 AND bp$\_$rp$>$=0.5}\\
{\it   AND parallax $<$= 3 AND i$\_$psf$>$=9)}
\\

We ran this query at ASVO\footnote{http://skymappertap.asvo.nci.org.au/ncitap/tap} with and without the condition $parallax ~>$=0.1.
The colour 
absolute magnitude diagram (CMD) (Figure 1) was 
changed only at the few percent level, 
suggesting that statistical biases in the absolute magnitudes
shown here may be 
modest.   
The rather flat distribution of parallaxes, 
with only a few percent negative parallaxes (Figure 2) in our CMD
is brought about  by the apparent magnitude cut in $G$ and the sparsity of stars in
the outer halo. Practical guards against SkyMapper photometric saturation include the NIMA flag 
the number of flagged pixels from bad, saturated and crosstalk pixel masks) and the
absence in Table 1 of stars unexpectedly red in $i-z$. 
Obviously, stars with apparently negative parallaxes have no place in Figure 1. The 3 mas cut is to control against thin disk stars, a population without a clear TRGB.

Modifications to the query, changing the magnitude limit from 14 to 14.5, and the latitude limit from --36 to --32 bring in a few times more stars, but do not seriously change
what we report in the next section. We also checked the $phot\_bp\_rp\_excess\_factor$
and found it averaged 1.2 $\pm$ 0.1, close to the no-contamination envelope
(Evans et al 2018).
Deviant stars could be rejected from the sample. 
Individual reddenings can be included by appending $ebmv\_sfd$ to the query list. These values are from Schlegel et al (1998).
\subsection{Simulations}
Measurement of luminosities from fluxes and parallaxes is nevertheless subject to the biases discussed by Eddington (1913), 
Lutz \& Kelker  (1975) and Arenou et al. (2018). 
Malmquist(1922) offered an analytic solution to the absolute magnitude bias
resulting in an apparent magnitude limited sample of uniform density
due to the larger volume accessible to the observer for brighter stars drawn from a distribution
of standard candle stars with a finite dispersion. Lutz \& Kelker addressed
the effect of parallax errors in a similar way. 
The question arises as to the dispersion of the TRGB across different stellar populations and mixed stellar populations, such as that of the thick disk and halo. This has been discussed in the review by Beaton et al (2018) of distance indicators for old stellar populations. According to Serenelli et al (2018) in the interval 10$^{-4}~ < ~Z~ < 4 \times 10^{-3}$ M$_{bol}$ varies by just 0.02 mag over an age range from 8.5 to 13.5 Gyrs. According to Sweigart \& Gross (1978) in the (0.7, 0.9) M$_\odot$ total mass range M$_{bol}$ varies by $\sim$0.15 mag for a 0.1 change in the helium abundance, Y. Such a large helium spread in the halo cannot be ruled out (e.g. Milone et al 2018). It would be appropriate to include intrinsic dispersion in the TRGB of this order in Malmquist bias simulations, and also to include the asymptotic giant branch which tends to erode the TRGB step function in the luminosity function of real stellar populations. Variations in metallicity, on the other hand, are mapped into colour variations. Fitting the TRGB to the CMD rather than the luminosity function avoids systematic errors and Malmquist bias due to the metallicity variable. If half of all field stars are binaries, this too might be considered in estimating intrinsic dispersion. Unless the components are identical in initial mass, however, the lower mass star will still be on the main sequence and effectively invisible, when the higher mass component reaches the helium flash.

One dimensional simulations with a power law luminosity function
terminating in the TRGB, a power law distribution of stellar distances starting\footnote{A suitable starting point for a power law in distance from the Sun, chosen to represent the halo. The choice does not qualitatively affect the results.} at 300 pc, a limiting flux, and a gaussian
distribution of $\sigma/\varpi$ whose mean absolute value is similar to that of the sample show that 
a quarter of the stars within 0.75 mag of the TRGB can be scattered into a region up to 0.75 mag brighter than the TRGB. Detailed 3D simulations with Galactic structure priors and an asymptotic giant branch 
are required to accurately correct biases of this kind.
We explore this in an Appendix, using the Besancon model of
stellar population synthesis (Robin et al 2003). Like Malmquist bias, we expect
corrections to M$_{TRGB}$ to scale as $\sigma^2$. The reduction in $\sigma$
in future Gaia data releases will markedly reduce the bias correction.

\section{Analysis}
\subsection{The brightest stars}
For stars with absolute magnitude $<$ 0, Figure 3 shows the distribution of our sample in celestial coordinates, centred on the South Galactic Pole (SGP). Gaia DR2 mean parallax
offsets are discussed by Arenou et al (2018) and Chen et al (2018). Adopting a
single mean parallax offset for the entire sample 
of --0.028 $\pm$ 0.006 mas (from Table 1 of Arenou et al. 2018), we collected
stars with M$_I~<$ --3.8 mag in Table 1.

No reddening corrections have been made to our sample, as high latitude extinction
is low (Burstein \& Heiles 1982; Schlegel et al 1998). 
The average extinction of stars in Table 1 is $A_I$ = 0.06 $\pm$ 0.002 mag.

\begin{table*}[h]
\centering
\label{my-label}
\caption{Brightest stars}
\begin{tabular}{rrrrccccrrc}
\hline
\textbf{\#} & \textbf{NIMA}&RA &(2000) Dec&G$_{BP}$--G$_{RP}$&$\varpi$
&$\sigma$&M$_I$&$i_{PSF}$&$z_{PSF}$&axn \\ 
&&(deg)&(deg)&(mag)&(mas)&(mas)&(mag)&(mag)&(mag)\\
\hline
  7582&      9&  11.3834& -37.3116&     2.14&     0.20&     0.04&    -3.94&    10.00&     9.69&     0.00\\
 26859&     12&  43.5734& -39.1784&     1.69&     0.22&     0.02&    -3.81&     9.94&     9.76&     0.00\\
 42247&      9&  17.6479& -60.3328&     1.58&     0.20&     0.02&    -3.96&    10.02&     9.85&     0.00\\
 45991&      0&  10.7259& -56.4505&     1.74&     0.14&     0.03&    -3.96&    10.73&    10.47&     0.00\\
 80820&      0&  36.7818& -60.3479&     1.84&     0.09&     0.02&    -4.25&    11.31&    11.06&     0.00\\
 82724&      7&  28.2021& -67.2882&     1.62&     0.19&     0.02&    -3.93&    10.05&     9.99&     0.00\\
 88959&      0&  61.1176& -40.3150&     1.78&     0.09&     0.02&    -4.07&    11.55&    11.26&     0.00\\
104253&      2&   5.7806& -72.0082&     1.91&     0.19&     0.02&    -4.06&     9.96&     9.66&     0.00\\
105485&      2&   6.0936& -71.8913&     2.39&     0.15&     0.03&    -4.50&    10.06&     9.66&     0.09\\
105788&      1&   5.6494& -72.1865&     2.45&     0.20&     0.04&    -3.87&    10.06&     9.57&     0.10\\
106058&      0&   5.5745& -72.1034&     2.01&     0.17&     0.02&    -3.97&    10.30&     9.98&     0.00\\
108127&      0& 324.4068& -48.3893&     1.78&     0.18&     0.04&    -4.36&     9.83&     9.57&     0.00\\
112897&      0&  49.9148& -69.3036&     2.00&     0.16&     0.02&    -4.03&    10.35&    10.02&     0.00\\
115208&      0&  12.1021& -71.6952&     1.80&     0.11&     0.02&    -3.83&    11.46&    11.20&     0.00\\
115856&      0&  15.8410& -70.9056&     2.15&     0.11&     0.03&    -4.05&    11.13&    10.89&     0.00\\
115997&     11&  34.6207& -68.3519&     1.74&     0.20&     0.02&    -4.00&     9.88&     9.82&     0.00\\
117828&      0&  41.0410& -69.9399&     1.49&     0.16&     0.02&    -3.86&    10.58&    10.40&     0.00\\
126917&      0&  65.9157& -50.5347&     1.82&     0.10&     0.02&    -4.02&    11.41&    11.11&     0.00\\
128085&      0&  65.8571& -37.4805&     1.80&     0.06&     0.01&    -3.92&    12.53&    12.29&     0.05\\
135000&      0& 336.8902& -71.3710&     2.01&     0.15&     0.04&    -4.17&    10.40&    10.03&     0.10\\
135776&      5&  71.2500& -37.3669&     2.10&     0.13&     0.02&    -4.34&    10.48&    10.04&     0.00\\
145140&      0&   6.9414& -75.0042&     1.52&     0.06&     0.01&    -3.99&    12.73&    12.54&     0.00\\
146835&     14& 326.1321& -71.0173&     1.83&     0.27&     0.03&    -4.03&     9.26&     9.06&     0.00\\
149332&      0&  69.6974& -60.7894&     2.43&     0.14&     0.02&    -3.88&    10.84&    10.33&     0.00\\
156327&      0&  40.6427& -73.3417&     1.45&     0.10&     0.02&    -3.91&    11.55&    11.37&     0.00\\
170142&      0&  72.4314& -54.1730&     1.86&     0.12&     0.03&    -4.08&    10.96&    10.66&     0.00\\
172993&      0&  73.4521& -28.6192&     1.99&     0.21&     0.03&    -3.85&     9.96&     9.59&     0.00\\
181392&      1& 312.1366& -59.3770&     1.99&     0.15&     0.03&    -4.24&    10.38&    10.05&     0.06\\
182005&      0&  73.3274& -63.0941&     1.88&     0.12&     0.03&    -4.37&    10.64&    10.36&     0.00\\
189778&      0&  75.1001& -35.3876&     1.82&     0.09&     0.02&    -4.51&    11.04&    10.74&     0.00\\
191603&     14&  68.3708& -12.5584&     1.80&     0.20&     0.03&    -4.16&     9.75&     9.57&     0.00\\
193812&     13& 309.3248& -35.5872&     1.60&     0.20&     0.04&    -4.06&     9.88&     9.51&     0.00\\
197291&      0& 308.3549& -50.7805&     1.49&     0.18&     0.04&    -3.87&    10.25&    10.08&     0.00\\
200297&      0&  71.8795& -65.3306&     1.52&     0.18&     0.03&    -4.00&    10.20&     9.98&     0.00\\

\hline

\multicolumn{11}{l}{$Notes$: NIMA is the \# of SkyMapper $i$ flagged pixels, i$\_$nimaflags, in the query}\\
\multicolumn{11}{l}{axn is the astrometric excess noise in the Gaia DR2 database}\\
\multicolumn{11}{l}{$\pi$ is the offset corrected parallax and $\sigma$ is parallax uncertainty}\\
\multicolumn{11}{l}{M$_I$ is the absolute I magnitude on the Cousins system.}\\
\multicolumn{11}{l}{No corrections for interstellar absorption have been applied to the SkyMapper 
mags.}
\end{tabular}
\end{table*}

\begin{figure*}
\begin{center}
\includegraphics[width=1.5\columnwidth,angle=-90]{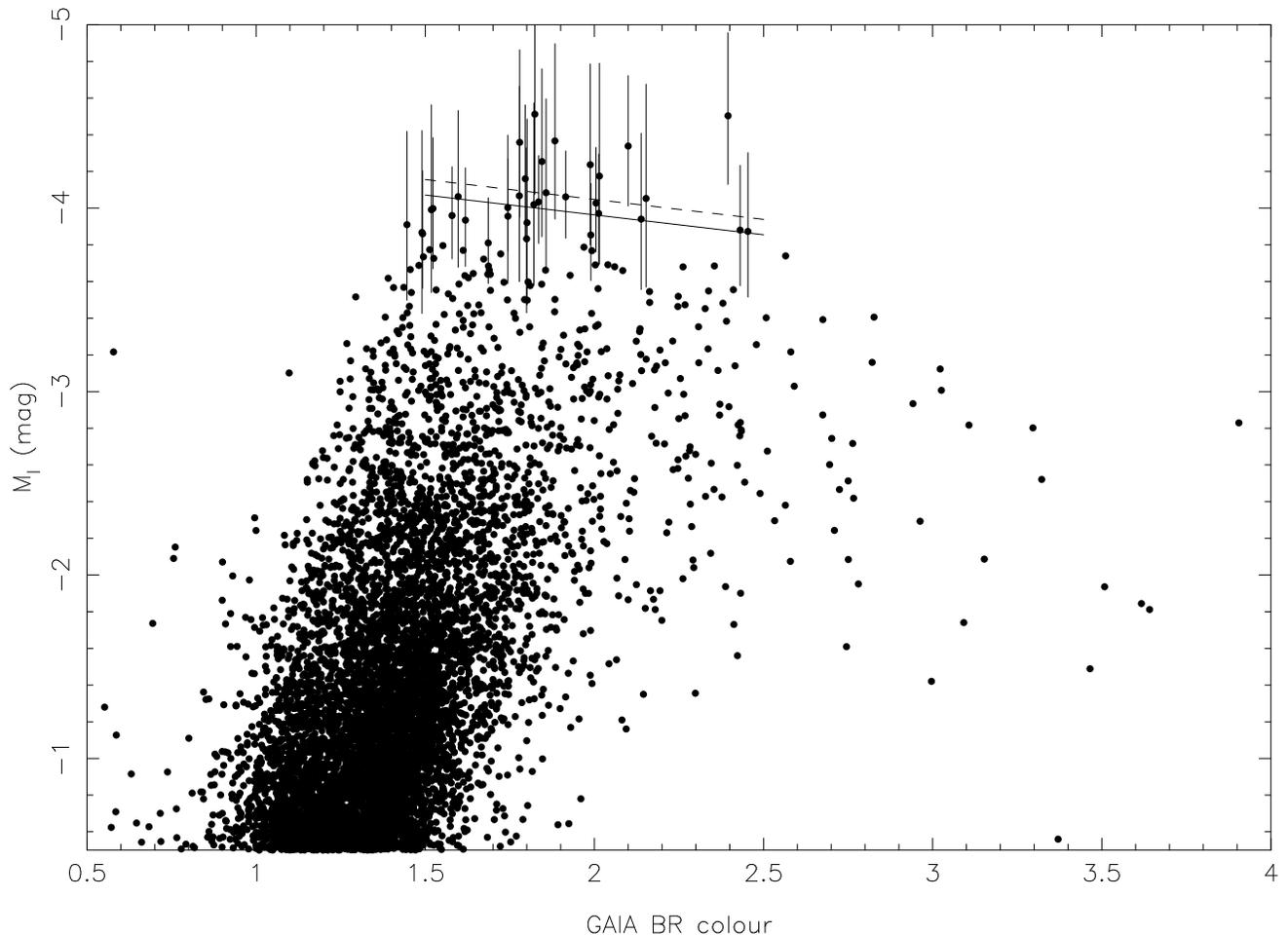}
\caption{SkyMapper Gaia DR2 red giant branch. The axis labelled BR colour is G$_{BP}$ -- G$_{RP}$. Parallax 
uncertainties are shown for the brightest stars.
A calibration of the TRGB by Rizzi et al (2007) is the solid line. A Gaia DR2 mean parallax offset of --0.028 $\pm$ 0.006 mas was 
subtracted from the data from Table 1 of Arenou et al (2018),
 appropriate to the Sculptor galaxy, which is at the South Galactic Pole. The dashed line represents its 6 $\mu$as uncertainty. 
The query in $\S$2.1 also 
brings in stars with luminosities fainter than are plotted here.}
\end{center}
\end{figure*} 

\begin{figure*}
\begin{center}
\includegraphics[width=\columnwidth,angle=-90]{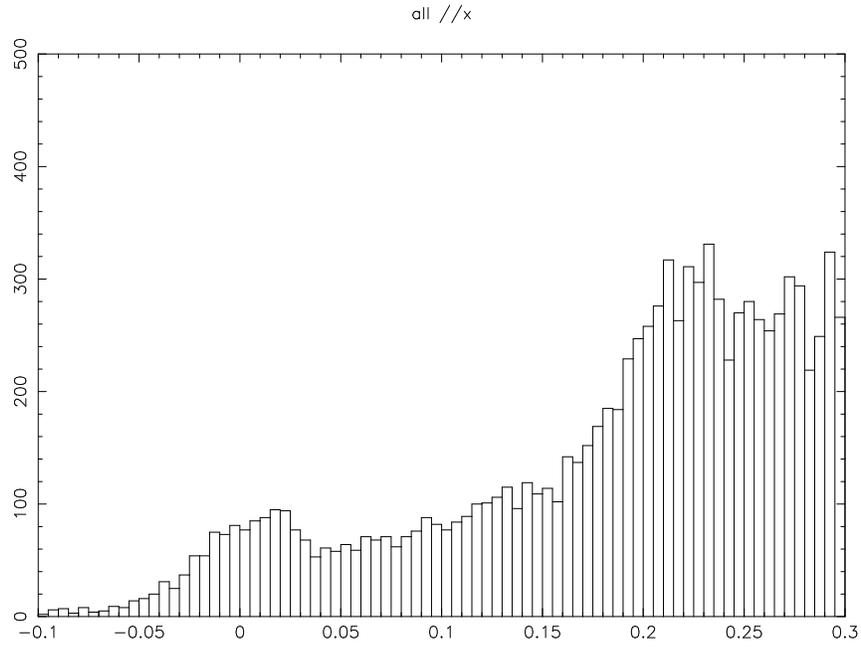}
\caption{
All parallaxes delivered by the query in mas units.}
\end{center}
\end{figure*} 
\begin{figure*}
\begin{center}
\includegraphics[width=1.25\columnwidth,angle=-90]{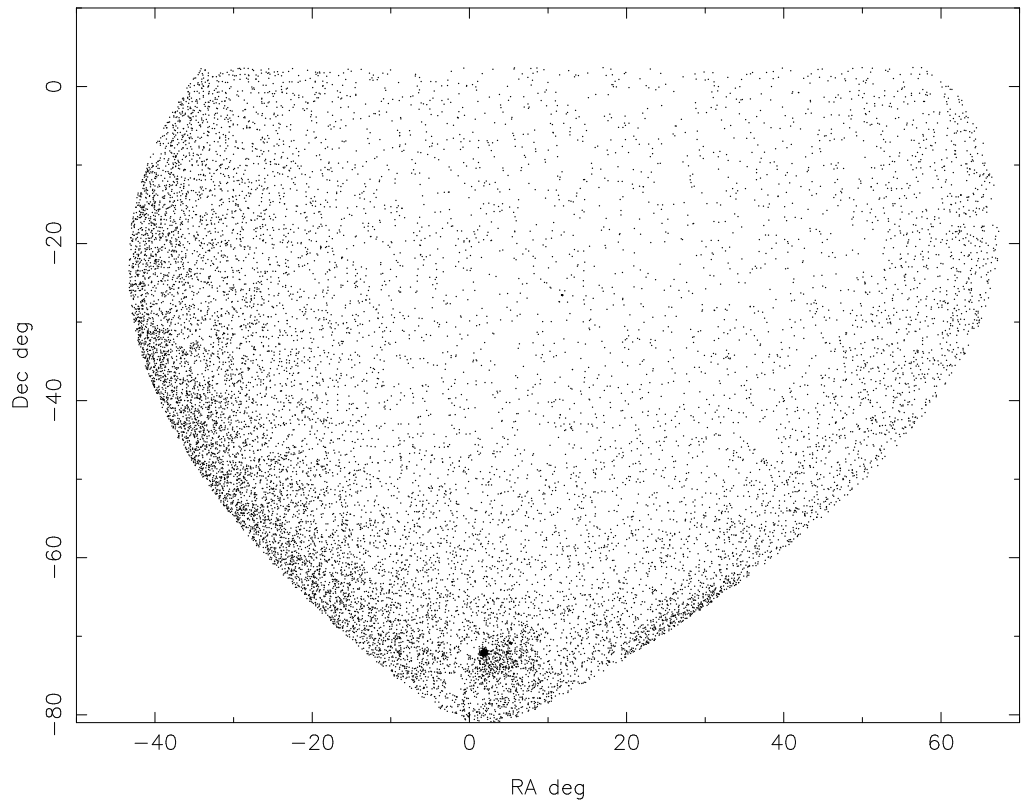}
\caption{Distribution of stars in Figure 1. The Small Magellanic Cloud and the cluster 47 Tuc can be seen at $\delta~\sim$ --72$^\circ$.}
\end{center}
\end{figure*} 
\subsection{Calibration}
In population II stars with a fixed helium abundance the TRGB luminosity is predicted to be a slow rising function of metallicity (Brown et al 2005). The bolometric correction is a function of effective temperature, but such temperatures are difficult to establish given the extensive convective envelopes of red giant stars. So the calibration of the TRGB has tended to be an empirical matter employing filters, such as $I$, $i$ and $z$, whose bandpasses include the peak of the Planck curve for red giants thereby reducing colour and metallicity dependence. 

Bellazzini (2007) finds $$M_I^{TRGB} = 0.08 (V-I)_0^2 - 0.194 (V-I)_0 - 3.939$$ on the Cousins photometric
system. Rizzi et al (2007) give $$M_I^{TRGB} = 0.215 [(V-I)_0 - 1.6] - 4.05$$  which,
following Hislop et al (2011), we adopt. It is, however, necessary to convert the AB-mag based SkyMapper $i$ magnitudes to the Vega-mag based Cousins I magnitudes.  We have investigated this using photometry for the Galactic globular clusters NGC 4590 (M68) and NGC 104 (47 Tuc). Cousins V and I magnitudes for RGB stars were selected from the photometric standard star fields data base maintained by Stetson\footnote{http://www3.cadc-ccda.hia-iha.nrc-cnrc.gc.ca/community/STETSON/
standards/} and cross-matched with both SkyMapper and Gaia DR2.  Reddening correction assumed the E(B-V) values for each cluster as tabulated in the current on-line version of the Harris (1996) catalogue.  In this process we assumed E(G$_{BP}$ -- G$_{RP}$)
 = E(V-I) based on the effective wavelength of the filter 
responses\footnote{
Gaia collaboration 
 et al (2018a) give a polynomial whose linear term is 1.04 
at the stellar colour and extinction relevant here versus the value of unity we have adopted.}. 
 Figure 4  then shows that for stars with G$_{BP}$ -- G$_{RP}$ 
$<$ 2, i.e., those where the difference in filter bandpasses does not show influence from the onset of TiO bands, is 0.435 $\pm$ 0.02 mag.  We have applied this offset to the SkyMapper $i$ mags to convert them to I mags on the Cousins system. 
For the same set of stars we have also investigated the relation between (V-I)$_0$ (Cousins system) and the dereddened (G$_{BP}$ -- G$_{RP}$)
colours derived from Gaia DR2.  As shown in Figure 5 a linear relation (G$_{BP}$ -- G$_{RP})_0$ 
$~=~~ (V-I)_0$ + 0.15 is an adequate representation\footnote{
http://gea.esac.esa.int/archive/documentation/GDR2/Data$\_$processing/
chap$\_$cu5pho/sec$\_$calibr/ssec$\_$cu5pho$\_$PhotTransf.html gives a quadratic relation, 
whose slope d(G$_{BP}$ -- G$_{RP}$)/d(V--I) is 0.1 mag larger at V--I = 1 than our adopted fit.}
of the relation. Figures 4 and 5 show the close relationship between Gaia colour and Cousins V--I.

\begin{figure}
\begin{center}
\includegraphics[width=.75\columnwidth,angle=-90]{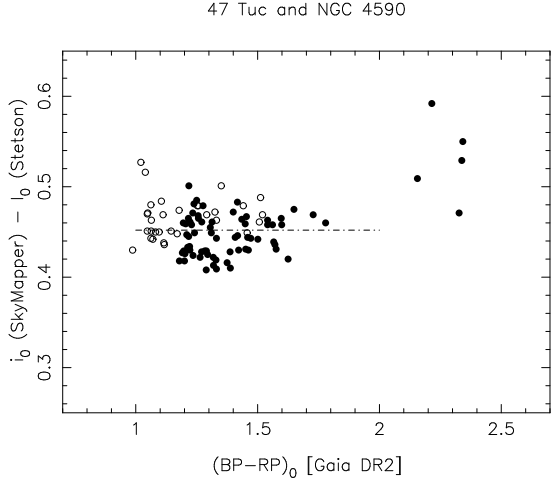}
\caption{For 99 stars with (G$_{BP}$ -- G$_{RP}$)
$<$ 2 mag, the difference between Cousins I magnitude and SkyMapper $i$ is 0.45 $\pm$ 0.02 mag, close to what is expected from the Vega vs AB zeropoints. In this and the following figure the axis labelled BP-RP is G$_{BP}$ -- G$_{RP}$}
\end{center}
\end{figure} 
\begin{figure}
\begin{center}
\includegraphics[width=.75\columnwidth,angle=-90]{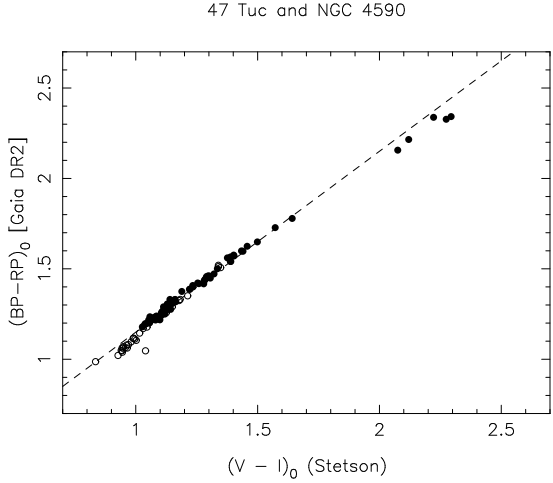}
\caption{On the Cousins V--I system (G$_{BP}$ -- G$_{RP})_0$  
$ ~=~ (V-I)_0$ + 0.15 is the dashed line.}
\end{center}
\end{figure} 
\subsection{Working in parallax space}
A proven solution to the biases in the astronomical distance scale
that result from working in luminosity space is to invert the distance indicator or scaling relation.
For example, assumed distances can be used to predict periods in the Cepheid period luminosity relation.
These can then be compared with observed periods. This approach is recommended for Gaia data by Arenou et al. In the case of the TRGB there is no relation to invert, but it is still possible to work in parallax space. We calculate

$$\chi^2 =\sum \frac{w_i  (\varpi_i - \varpi (predicted))^2}{
(\sigma_i^2 + (\delta_i \varpi_i /2.16)^2)} (\Sigma w_i)^{-1} $$  
$${\rm with~}  w_i = 1 {~\rm for~} |M_i-TRGB|~<~\epsilon$$
$${\rm and~} w_i = 0 {~\rm for~} |M_i-TRGB|~>~\epsilon \eqno{(1)}$$

\noindent where $\varpi_i$ is the star's observed parallax $\varpi~(predicted)$ is the prediction from the TRGB
and $\sigma_i$ is the star's parallax error and $\delta_i$ its photometric error in magnitudes. 
We chose values of $\epsilon$ between 0.25 and 0.5 mag and obtained clear minima close to unity
in $\chi^2$ per degree of freedom.

To eliminate the arbitrariness of $\epsilon$ we can adopt $$w_i = exp^{-((M_i-TRGB)\varpi_i /\sigma_i/2.16)^2} \eqno{(2)}$$ to downweight stars of 
absolute magnitude M$_i$ in Figure 1 that are far from the TRGB.
With this weighting equation (1) is a figure of merit (or demerit), rather than a $\chi^2$ $per~se$.
Figure 
6 is a map of this $\chi^2$ in (TRGB, $\sigma/\varpi$) space where $\sigma/\varpi$ is the threshold applied
to parallax error for inclusion in the summation. Although the data are noisy with the present limited sample, there is evidently a minimum of $\chi^2$ in the vicinity of TRGB = --4 mag that is fairly independent of the $\sigma/\varpi$ cut.
\begin{figure}
\begin{center}
\includegraphics[width=\columnwidth,angle=-90]{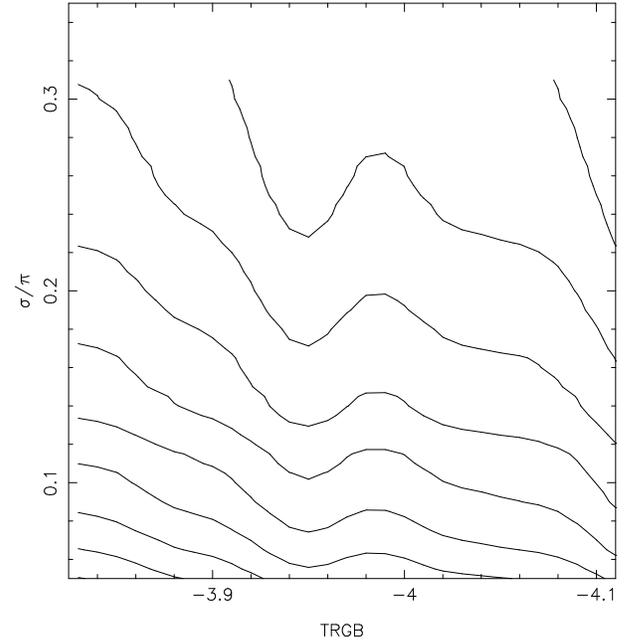}
\caption{Contours of $\chi^2$ mapped against TRGB and $\sigma/\varpi$ cut.
There is a clear local minimum between --3.92 and --3.98 independent of
parallax error cut.}
\end{center}
\end{figure}

Bootstrap resampling can be employed to obtain a confidence interval for TRGB~at the fiducial colour\footnote{Hislop et al (2011).} of V--I = 1.6. It is evident from Figure 6 that cutting the sample in two in $\sigma/\varpi$ affects the result less than the uncertainty in the parallax offset does. 
This yields an
uncertainty in M$_I$ for TRGB of 0.076 mag.
We explored the effect of the weights in equation (2) 
by offsetting the input
magnitudes with a small blind offset and found the local minimum in Figure 6 moved as expected. 

\section{Summary and future work}
We find that a high latitude Gaia DR2 CMD is consistent with the current calibration of the TRGB as an extragalactic distance indicator. This can be improved as follows
\begin{itemize}
\item detailed simulations of parallax uncertainty luminosity bias
\item adding PanSTARRS photometry of high northern latitudes\footnote{The ESA Gaia archive
includes this matched catalog. With the aid of the helpdesk we used a query similar to that in $\S$2.1
to obtain a CMD similar to Figure 1 for the northern galactic hemisphere. Wolf et al (2018) give a simple $i$ band photometric transformation between SkyMapper and PanSTARRS.}
\item reducing the uncertainty in parallaxes and the parallax offset in future Gaia data releases
\item extending the wavelength range through infrared photometry of stars with M$_I~<$ --3.8 mag.
\end{itemize}

This work has made use of data from the ESA space mission Gaia, processed by the Gaia Data Processing and Analysis Consortium (DPAC). Funding for the DPAC has been provided by national institutions participating  in the Gaia Multilateral Agreement.
The national facility capability for SkyMapper has been funded through ARC LIEF grant LE130100104 from the Australian Research Council, awarded to the University of Sydney, the Australian National University, Swinburne University of Technology, the University of Queensland, the University of Western Australia, the University of Melbourne, Curtin University of Technology, Monash University and the Australian Astronomical Observatory. SkyMapper is owned and operated by The Australian National University's Research School of Astronomy and Astrophysics. The survey data were processed and provided by the SkyMapper Team at ANU. The SkyMapper node of the All-Sky Virtual Observatory (ASVO) is hosted at the National Computational Infrastructure (NCI). Development and support the SkyMapper node of the ASVO has been funded in part by Astronomy Australia Limited (AAL) and the Australian Government through the Commonwealth's Education Investment Fund (EIF) and National Collaborative Research Infrastructure Strategy (NCRIS), particularly the National eResearch Collaboration Tools and Resources (NeCTAR) and the Australian National Data Service Projects (ANDS). Parts of this project were conducted by the Australian Research Council Centre of Excellence for All-sky Astrophysics (CAASTRO), through project number CE110001020. Two of us are grateful for the hospitality of MIAPP, where the first draft of this paper was made and to the organizers of MIAPP's workshop on the Extragalactic Distance Scale in the Gaia Era, Rolf Kudritzki, Lucas Macri and Sherry Suyu. This research was supported by the Munich Institute for Astro- and Particle Physics (MIAPP) of the DFG cluster of excellence ``Origin and Structure of the Universe''. We thank Christian Wolf and Chris Onken for their help formulating and implementing the database query.
We thank an anonymous referee for comments that improved the paper.

\section*{References}
Arenou, F. et al 2018, A\&A, 616, A17\\
Gaia Collaboration, Babusiaux, C., et al 2018a, A\&A, 616, A10\\
Beaton, R. et al 2016, ApJ, 832, 210\\
Beaton, R. et al 2018, arxiv 180889191\\
Bellazzini, M. 2007, MemSAIt, 79, 440\\
Gaia Collaboration, Brown, A.G.A. et al 2018b, A\&A 616, A1\\
Brown, T. et al 2005, AJ, 130, 1693\\
Burstein, D. \& Heiles, C. 1982, AJ, 87, 1165\\
Chen, S. et al 2018, arxiv 1807.07089\\
Clementini, G. et al 2018, A\&A (in press), arxiv 1805.02079\\
Da Costa, G. \& Armandroff, T. 1990, AJ, 100, 162\\
Di Valentino, E. et al 2018, JCAP 04, 017\\
Eddington, A. 1913, MNRAS, 73, 359\\
Evans, D. et al 2018, A\&A 616, A4\\
Harris, W. E. 1996, AJ, 112, 1487
\\
Hatt, D., et al., 2018, ApJ, 861, 104\\
Hernitschek, N., et al., 2018, ApJ, 859, 31\\
Hislop, L. et al 2011, ApJ, 733, 75\\
Iorio, G. et al 
 2018, MNRAS, 474, 2142\\
Luri, X. et al 2018, A\&A 616, A9\\
Lutz, T. \& Kelker, D. 1975, PASP, 85, 574\\
Malmquist, K. 1922, Meddelanden Lund Obs., 100, 1\\
Milone, A. et al 2018, MNRAS, 481, 5098\\
Mould, J. 2017, Nature Astronomy, 1, 739\\
Rizzi, L. et al 2007, ApJ, 661, 815\\
Robin, A. et al 2003, A\&A, 409, 523\\
Saha, A. 1985, ApJ, 289, 310\\
Schlegel, D. et al 1998, ApJ, 500, 525\\
Serenelli, A. et al 2018, A\&A, 606, 33\\
Sweigart, A. \& Gross, P. 1978, ApJS, 36, 405\\
Wolf, C. et al 2018, PASA, 35, 010

\section*{Appendix}

\setcounter{figure}{0}

Here we outline the parameters used to create the Besancon model (Robin et al.\ 2003) that has been used to explore potential biases in the observed luminosity distribution.
The model was generated to broadly resemble the underlying population from which the observed sample shown in Figure 1 was selected.
The specific parameters adopted for the model were:
\begin{enumerate}
\item --180 $\leq$ {\it l} $\leq$ +180 with steps of 30 deg;
\item 30 $\leq$ |{\it b}| $\leq$ 90 with steps of 30 deg;
\item distances between 0.3 and 50 kpc;
\item all absolute magnitudes (i.e., the default --7 $\leq$ M$_{V}$ $\leq$ 20);
\item all stellar populations (i.e., all ages from young disk (0. -- 0.15 Gyr) through to old disk (7 -- 10 Gyr)), thick disk, halo and bulge)
with the default range of metallicities for each population;
\item spectral types from G0 to M5 with all luminosity classes included;
\item stars selected to have 10 $\leq$ $V$ $\leq$ 16.
\end{enumerate}
Application of these parameters then resulted in a model containing 1.41 million stars.  In Figure A1 we show the $M_{I}, (V-I)_{0}$ colour-magnitude diagram, on the Cousins system, for the 44847 stars in the model that are brighter than $M_{I}$ = --0.5 mag.  While the effects of the discrete sampling of the model are obvious, it is evident that the most luminous stars in the model are predominantly metal-poor halo RGB stars.
\begin{figure}
\begin{center}
\includegraphics[width=0.56\columnwidth,angle=-90]{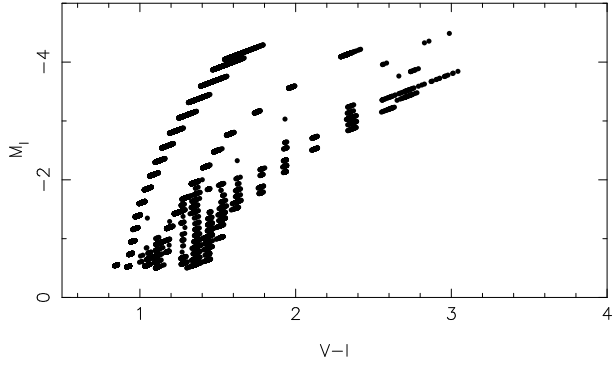}
\caption{$M_{I}, (V-I)_{0}$ colour-magnitude diagram for the 44847 stars in the generated Besancon model that more luminous than $M_{I}$ = --0.5 mag.  Note the predominance of the halo RGB at the brightest luminosities.}
\end{center}
\end{figure}
Figure A2 shows in black the $M_{I}$ luminosity function (LF) for the stars in Figure A1 more luminous than $M_{I}$ = --1.5.  We note that the discrete nature of the model generation process, which fundamentally uses $V$ magnitudes, results in an uneven LF for the 0.1 mag bins used in generating the histogram -- a broader binning yields a smoother function.  Shown also in the Figure are the results of convolving the model LF with a median parallax error amplitudes of 0.019 mas (green line) and 0.038 mas (red line). (The median parallax error
is almost independent of magnitude for 11 $<~i~<$ 9.) This confirms that parallax errors can scatter stars near the RGB tip by up to $\sim$1 mag brighter. 

A control on this bias is available by making cuts in $\sigma/\varpi$.
The lower part of Figure A2 shows the effect of cuts $\sigma/\varpi~<$
0.2, 0.4. Lutz \& Kelker (1975) found a similar effect with subdwarfs.

\begin{figure}
\begin{center}
\includegraphics[width=1.5\columnwidth,angle=-90]{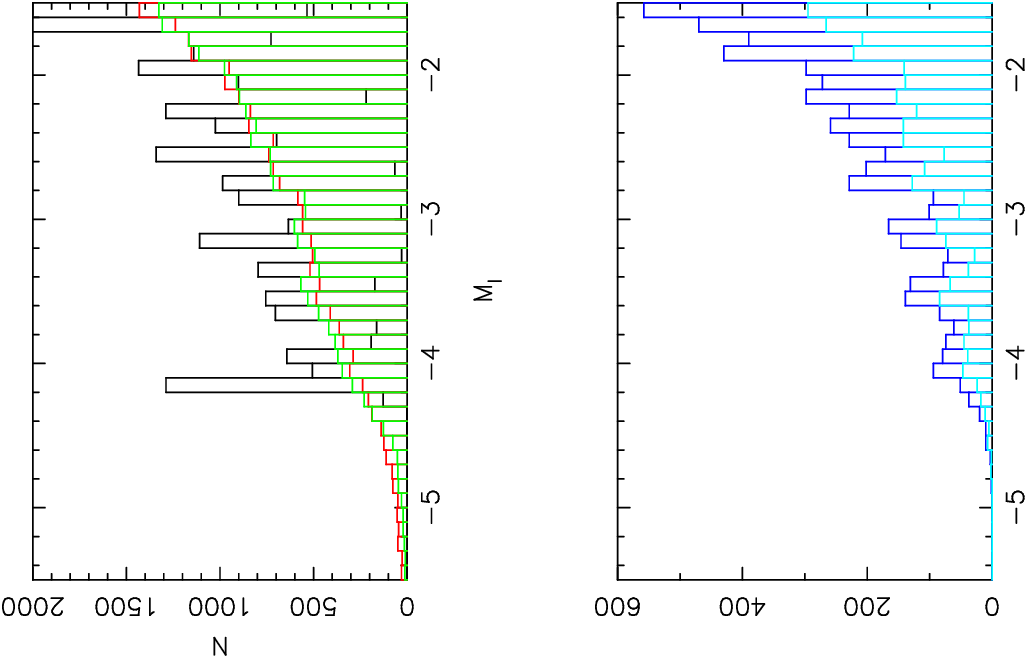}
\caption{Black: Besancon model luminosity function. Red: observed
with a uniform distribution of parallax error amplitude $\pm$0.02 mas.
Green: $\pm$0.01 mas.
The lower panel shows the effect of cuts in $\sigma/\varpi$.
Blue is $\sigma/\varpi~<$ 0.4. Light blue is $\sigma/\varpi~<$ 0.2,
the ``physics'' 5$\sigma$ cut.}
\end{center}
\end{figure}

\end{document}